\documentclass{article}[11pt]
\usepackage{a4wide}
\usepackage{bm}
\usepackage{bbm}
\usepackage{epsfig,graphics}
\usepackage{amsmath,amssymb,amsfonts}
\usepackage{color}
\usepackage{epstopdf}
\usepackage[%dvipdfm,
colorlinks=true,
linkcolor=black,
breaklinks=true,
urlcolor=blue,
citecolor=green]{hyperref}

\newcommand{\Amp}{\mathcal{A}}
\newcommand{\mO}{\mathcal{O}}

\begin{document}

\title{ How to reveal the exotic nature of the $P_c(4450)$ }

\author{Feng-Kun Guo$^{1,2,}$\footnote{Email address:
      \texttt{fkguo@hiskp.uni-bonn.de} },~
      Ulf-G. Mei\ss ner$^{2,3,}$\footnote{Email address:
      \texttt{meissner@hiskp.uni-bonn.de} },~
      Wei Wang$^{4,1,}$\footnote{Email address:
      \texttt{wei.wang@sjtu.edu.cn} } ~ and
      Zhi Yang$^{2,}$\footnote{Email address:
      \texttt{zhiyang@hiskp.uni-bonn.de} } \\
      {\it\small$^1$State Key Laboratory of Theoretical Physics, Institute of
      Theoretical Physics, }\\
      {\it\small Chinese Academy of Science, Beijing 100190, China }\\
      {\it\small$^2$Helmholtz-Institut f\"ur Strahlen- und Kernphysik and Bethe
      Center for Theoretical Physics,}\\
      {\it\small Universit\"at Bonn, D-53115 Bonn, Germany}\\
      {\it\small$^3$Institute for Advanced Simulation, Institut f\"{u}r
       Kernphysik and J\"ulich Center for Hadron Physics,}\\
      {\it\small Forschungszentrum J\"{u}lich, D-52425 J\"{u}lich, Germany} \\
      {\it\small$^4$ INPAC, Shanghai Key Laboratory for Particle Physics and
      Cosmology, }\\
      {\it\small Department of Physics and Astronomy, Shanghai Jiao-Tong University, Shanghai 200240,   China}\\
       }
\date{\today}

\maketitle

\begin{abstract}

The LHCb Collaboration announced two pentaquark-like structures in the $J/\psi\,
p$ invariant mass distribution. We show that the current information on the
narrow structure at 4.45~GeV is compatible with kinematical effects of the
rescattering from $\chi_{c1}\, p$ to $J/\psi\, p$:
First, it is located exactly at the $\chi_{c1}\,p$ threshold. Second, the mass
of the four-star well-established $\Lambda(1890)$ is such that a leading Landau
singularity from a triangle diagram can coincidentally appear at the
$\chi_{c1}\,p$ threshold, and third, there is a narrow structure at the
$\chi_{c1}\,p$ threshold but not at the $\chi_{c0}\,p$ and $\chi_{c2}\,p$
thresholds. In order to check whether that structure corresponds to a real
exotic resonance, one can measure the process $\Lambda_b^0\to K^-\chi_{c1}\,p$.
If the $P_c(4450)$ structure exists in the $\chi_{c1}\,p$ invariant mass
distribution as well, then the structure cannot be just a kinematical effect but
is a real resonance, otherwise, one cannot conclude the $P_c(4450)$ to be
another exotic hadron. In addition, it is also worthwhile to measure the decay
$\Upsilon(1S)\to J/\psi\, p\, \bar p$: a narrow structure at 4.45~GeV but not at
the $\chi_{c0}\, p$ and $\chi_{c2}\, p$ thresholds would exclude the possibility
of a pure kinematical effect.

\end{abstract}

\medskip

\newpage

The observation of many different hadrons half a century ago stimulated the
proposal of the quark model as a classification scheme~\cite{GellMann:1964nj},
and helped to establish  quantum chromodynamics (QCD) as the fundamental theory
of the  strong interactions. Since then, hundreds of more hadrons were
discovered.
A renaissance of hadron spectroscopy studies started in  2003, and since then a
central topic is the identification of the so called exotic hadrons.
These are states  beyond the naive quark model scheme, in which  mesons and
baryons are composed of a quark--antiquark pair and three quarks, respectively.
Most of the new interesting structures were observed in the mass region of heavy
quarkonium, and are called  $XYZ$ states (for a list of these particles and a
review up to 2014, see Ref.~\cite{Agashe:2014kda}). In particular, the
$X(3872)$~\cite{Choi:2003ue} extremely close to the $D^0\bar D^{*0}$ threshold
is widely regarded as an exotic meson, and the charged structures with a hidden
pair of heavy quark and heavy antiquark such as the
$Z_c(4430)$~\cite{Choi:2007wga,Aaij:2014jqa},
$Z_c^\pm(3900)$~\cite{Ablikim:2013mio,Liu:2013dau},
$Z_c^\pm(4020)$~\cite{Ablikim:2013emm}, and
$Z_b^\pm(10610,10650)$~\cite{Belle:2011aa} would be explicitly exotic were they
really resonances, i.e. poles of the $S$-matrix. Candidates for explicitly
exotic hadrons were extended to the pentaquark sector by the new LHCb
observations of two structures, denoted as $P_c$,  in the $J/\psi\, p$ invariant
mass distribution with masses (widths) $(4380\pm8\pm29)$~MeV
($(205\pm18\pm86)$~MeV) and  $(4449.8\pm1.7\pm2.5)$~MeV ($(39\pm5\pm19)$~MeV),
respectively~\cite{Aaij:2015tga}. They were suggested to be hadronic molecules
composed of an anticharm meson and a charmed
baryon~\cite{Chen:2015loa,Chen:2015moa,Roca:2015dva} the existence of which were
already predicted in Refs.~\cite{Wu:2010jy,Wu:2010vk,Yang:2011wz,Xiao:2013yca}.
They were also discussed as a pentaquark doublet in Ref.~\cite{Mironov:2015ica}.

Normally, such structures are observed as peaks in invariant mass distributions
of certain final states, and fitted by using the Breit--Wigner parameterization
to extract the masses and widths.
However, such a procedure is problematic. On the one hand, many of these
structures are very close to certain thresholds to which they couple strongly.
In this case, a use of Breit--Wigner is questionable and one needs to account
for the thresholds. This can be achieved using the Flatt\'e
parameterization~\cite{Flatte:1976xu} (a method in this spirit for
near-threshold states with coupled channels and unitarity was recently proposed
in Ref.~\cite{Hanhart:2015cua}).
On the other hand, not every peak should be attributed to the existence of a
resonance. In particular, kinematical effects may also show up as peaks. Such
kinematical effects correspond to singularities of the $S$-matrix as well, but
they are not poles. In general, they are the so-called Landau singularities
including  branch points at thresholds and more complicated ones such as the
triangle singularity, also called anomalous threshold~(detailed discussions of
these singularities can be found in the textbooks~\cite{Eden:book,Chang:book}).
The observability of the triangle singularity was extensively discussed in
1960s (see Refs.~\cite{Schmid:1967,Aitchison:2015jxa,Liu:2015taa} and
references therein), and recently was used to explain some structures including
the $\eta(1405)$, $a_1(1420)$ and
$\phi(2170)$~\cite{Wu:2011yx,Wu:2012pg,Wang:2013hga,Ketzer:2015tqa,
Achasov:2015uua,Lorenz:2015pba}.
In fact, there were suggestions that some of the $Z_c$ and $Z_b$ states were
threshold effects~\cite{Rosner:2007mu,Bugg:2008wu,Bugg:2011jr,Swanson:2014tra,
Chen:2013coa} and the threshold effects might be enhanced by triangle
singularities~\cite{Szczepaniak:2015eza}.  For a general discussion of $S$-wave
threshold effects, see also Ref.~\cite{Rosner:2006vc}. Therefore, in order to
establish a structure as a resonance, one has to discriminate it from such
kinematical effects. Indeed, this is possible. As discussed in
Ref.~\cite{Guo:2014iya}, a resonance can be distinguished from threshold
kinematical effects only in the elastic channel which is the channel with that
threshold. The purpose of this paper is to discuss the possible kinematical
effects for the narrower structure at 4.45~GeV in the LHCb observations and
suggest measurements to check whether it is a real exotic resonance or not.

\begin{figure}[tbh]
  \centering
    \includegraphics[width=0.67\linewidth]{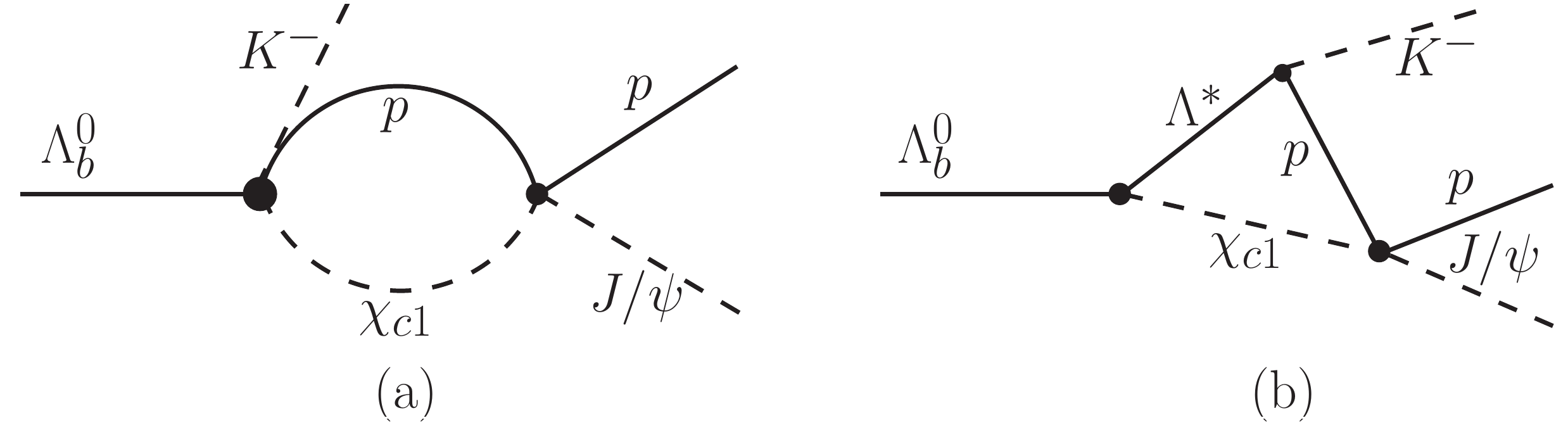}
  \caption{Two-point and three-point loops for the mechanism of
  the $\chi_{c1}\,p\to J/\psi\, p$ rescattering in the decay $\Lambda_b^0\to K^-
  J/\psi\, p$ .
  }
  \label{fig:diag}
\end{figure}

We first notice that the $P_c(4450)$ structure is exactly located at the
threshold of a pair of $\chi_{c1}$ and proton, $(4448.93\pm0.07)$~MeV, and
\begin{equation}
  M_{P_c(4050)} - M_{\chi_{c1}} - M_p = (0.9\pm3.1)~\text{MeV}.
\end{equation}
If the angular momentum between the $\chi_{c1}$ and proton is a $P$-wave, then the
two-body system can have quantum numbers $J^P=(1/2,\, 3/2,\, 5/2)^-$, compatible
with the favored possibilities $5/2^+,5/2^-$ and $3/2^-$~\cite{Aaij:2015tga}. The
$\chi_{c1}\,p$ can rescatter into the observed $J/\psi\, p$ by
exchanging soft gluons. Two possible diagrams for such a mechanism are shown in
Fig.~\ref{fig:diag}, where (a) is a two-point loop with a prompt three-body
production $\Lambda_b^0\to K^- \chi_{c1}\, p$ followed by the rescattering process
$\chi_{c1}\,p \to J/\psi\, p$, and in (b) the $K^- p$ pair is produced from an
intermediate $\Lambda^*$ state and the proton rescatters with the $\chi_{c1}$
into the $J/\psi\, p$. We will discuss them subsequently.
%%%%%%%%%%%%%%%%%%%%%%%%%%%%%%%%%%%%%%%%%%%%%%%%%%%%%%%%%%%%%%%%%%%%%%%%%%%%%%%%
%%%%%%%%%%%%%%%%%%%%%%%%%%%%%%%%%%%%%%%%%%%%%%%%%%%%%%%%%%%%%%%%%%%%%%%%%%%%%%%%

It is worthwhile to notice that the $\chi_{c1}$ can be produced in the weak
decays of the $\Lambda_b$ with a similar magnitude as that for the $J/\psi$.
In the bottom quark decays,  the charm quark is produced via the mediation of
the $W$-boson. After integrating out the off-shell mediators,  one arrives at
two effective operators  for the  $b\to c\bar c s$ transition:
\begin{eqnarray}
\mO_1= [\bar c^\alpha\gamma^\mu(1-\gamma_5) c^\alpha][ \bar
s^\beta\gamma_{\mu}(1-\gamma_5) b^\beta],\quad \mO_2= [\bar
c^{\alpha}\gamma^\mu(1-\gamma_5) c^{\beta}] [ \bar s^{\beta}\gamma_{\mu}(1-\gamma_5) b^{\alpha}]~,
\label{eq:operators}
\end{eqnarray}
where one-loop QCD corrections have been taken into account to form $\mO_1$.
Here, $\alpha, \beta$ are  color indices,  and they should be set to be the same in
 $\mO_2$ in order to form a color-singlet charmonium state.  The quark fields, $
 [\bar c\gamma^\mu(1-\gamma_5) c]$, will directly  generate the charmonium state.  A charmonium  with $J^{PC}=1^{--}$ like the $J/\psi$ is
produced by the vector current, while the axial-vector current tends to produce
the $\chi_{c1}$ with $J^{PC}=1^{++}$ and the $\eta_c$ state with $J^{PC}=0^{+-}$.
Since  the vector  and axial-vector currents have the same strength in the weak
operators, one would expect the production rates for the $J/\psi$ and $\chi_{c1}$
are of the same order in $b$ quark decays. Corrections to this expectation
come from  higher-order QCD contributions but are
sub-leading~\cite{Beneke:2008pi}. In fact, such an expectation is supported by
the $B$ meson decay data~\cite{Agashe:2014kda}:
\begin{eqnarray}
 {\cal B}(B^+\to J/\psi K^+)= (10.27\pm0.31)\times 10^{-4},\;\;\;
 {\cal B}(B^+\to \chi_{c1} K^+)= (4.79\pm0.23)\times 10^{-4}.
\end{eqnarray}

Having made these general observations, we return to the discussion of the
$\Lambda_b^0$ decays measured by LHCb.
We will first focus on the two-point loop diagram whose singularity is a branch
point at the $\chi_{c1}\,p$ threshold on the real axis of the complex $s$ plane,
where and in the following $\sqrt{s}$ denotes the invariant mass of the
$J/\psi\, p$ or $\chi_{c1}\,p$ system. It manifests itself as a cusp at the
threshold if the $\chi_{c1}\,p$ is in an $S$-wave. For higher partial waves, the
threshold behavior of the amplitude is more smooth and a cusp becomes evident in
derivatives of the amplitude with respect to $s$. Since we are only interested
in the near-threshold region, both of the $\chi_{c1}$ and the proton are
nonrelativistic. Thus, the amplitude for Fig.~\ref{fig:diag}~(a) is proportional
to the nonrelativistic two-point loop integral
\begin{equation}
G_\Lambda(E)=\int \frac{d^3 q}{(2\pi)^3}\frac{ \vec q\, ^2 \, f_\Lambda(\vec q\,
^2)}{E-m_1 -m_2 -\vec q\, ^2/(2\mu)} \ ,
\end{equation}
where $m_{1,2}$ denote the masses of the intermediate states in the loop, $\mu$
is the reduced mass and $E$ is the total energy. Here, we consider the case for
the $P$-wave $\chi_{c1}\,p$ which has quantum numbers compatible with the
possibilities of the $P_c(4450)$ reported by the LHCb Collaboration, though one
should be conservative to take these determinations for granted as
none of the singularities discussed here was taken into account in the LHCb
amplitude analysis.
If we take a Gaussian form factor, $f_\Lambda(\vec q\, ^2) =
\exp\left(-2\vec q\, ^2/\Lambda^2\right)$, to regularize the loop integral, the
analytic expression for the loop integral is then given by
\begin{eqnarray}
G_\Lambda(E) = -\frac{ \mu\, \Lambda }{(2\pi)^{3/2}} \left( k^2 +
\frac{\Lambda^2}{4} \right) + \frac{\mu\, k^3}{2\pi} e^{-2k^2/\Lambda^2} \left[
\text{erfi}\left(\frac{\sqrt{2}k}{\Lambda}\right) - i \right],
\label{eq:gexplicit}
\end{eqnarray}
with $k = \sqrt{2\mu (E-m_1-m_2 + i \epsilon)}$, and the imaginary error
function $\text{erfi}(z) = (2/{\sqrt{\pi}}) \int_0^z e^{t^2}dt$. A better
regularization method should be
applied in the future, but for our present   study such an approach is fine.

%%%%%%%%%%%%%%%%%%%%%%%%%%%%%%%%%%%%%%%%%%%%%%%%%%%%%%%%%%%%%%%%%%%%%%%%%%%%%%%%
\begin{figure}[tbh]
  \centering
    \includegraphics[width=0.6\linewidth]{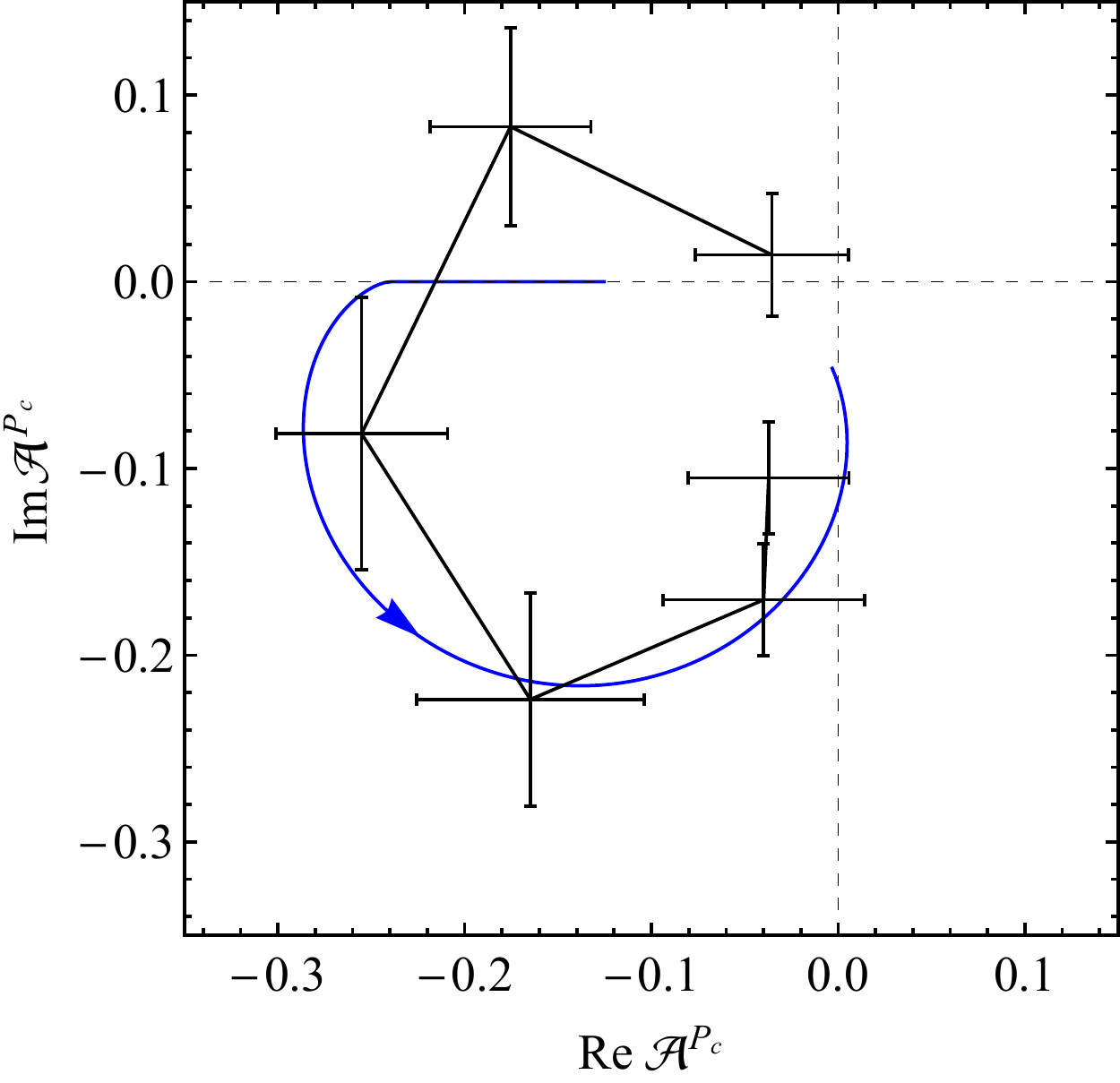}
  \caption{Fit to the real and imaginary parts of the $P_c(4450)$ amplitude
  shown in Fig.~9 in Ref.~\cite{Aaij:2015tga} with Eq.~\eqref{eq:Aa}. The blue
  curve represents the best fit. It is counterclockwise with increasing the
  $J\psi\, p$ invariant mass from 4.41~GeV to 4.49~GeV, the same range as for
  the LHCb diagram.
  }
  \label{fig:argand}
\end{figure}
%%%%%%%%%%%%%%%%%%%%%%%%%%%%%%%%%%%%%%%%%%%%%%%%%%%%%%%%%%%%%%%%%%%%%%%%%%%%%%%%

\begin{figure}[tb]
  \centering
    \includegraphics[width=0.49\linewidth]{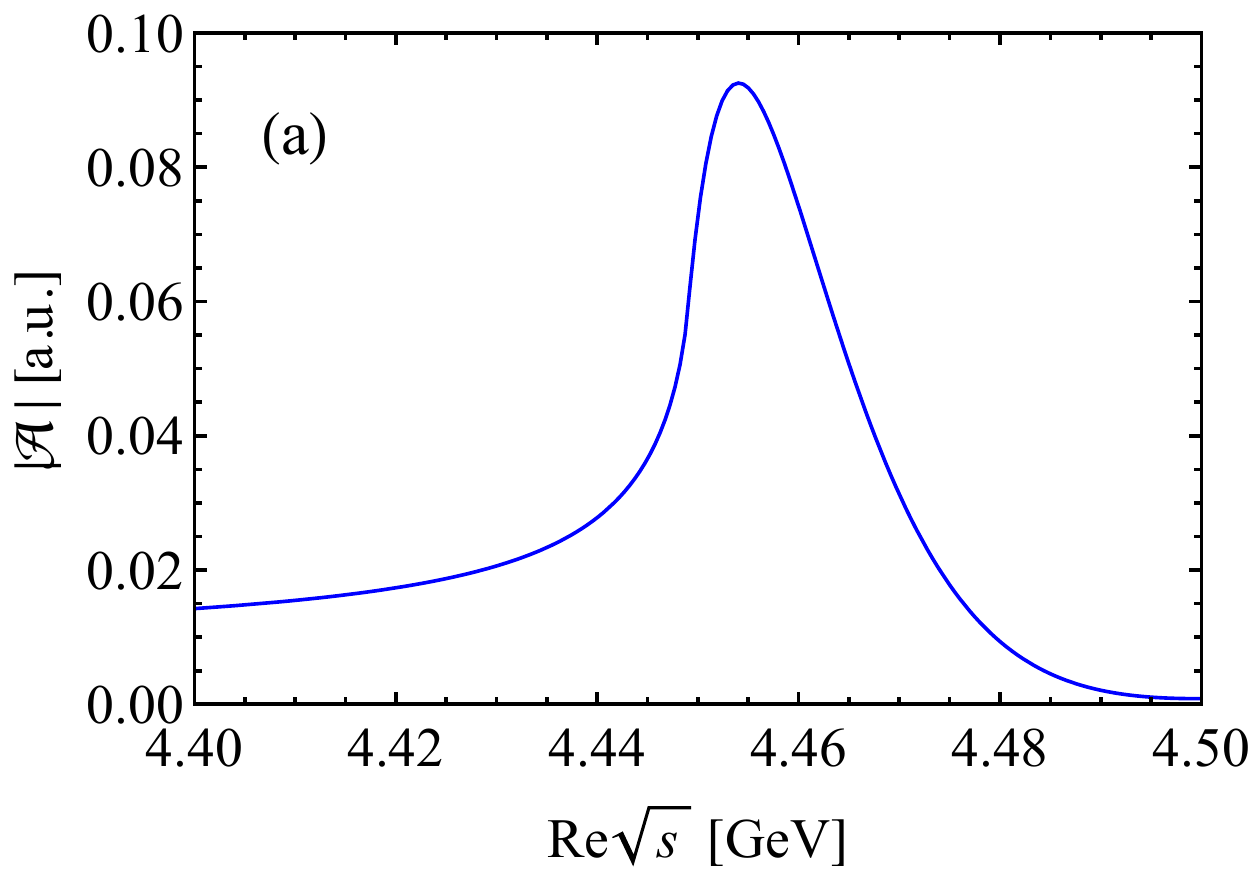} \hfill
    \includegraphics[width=0.49\linewidth]{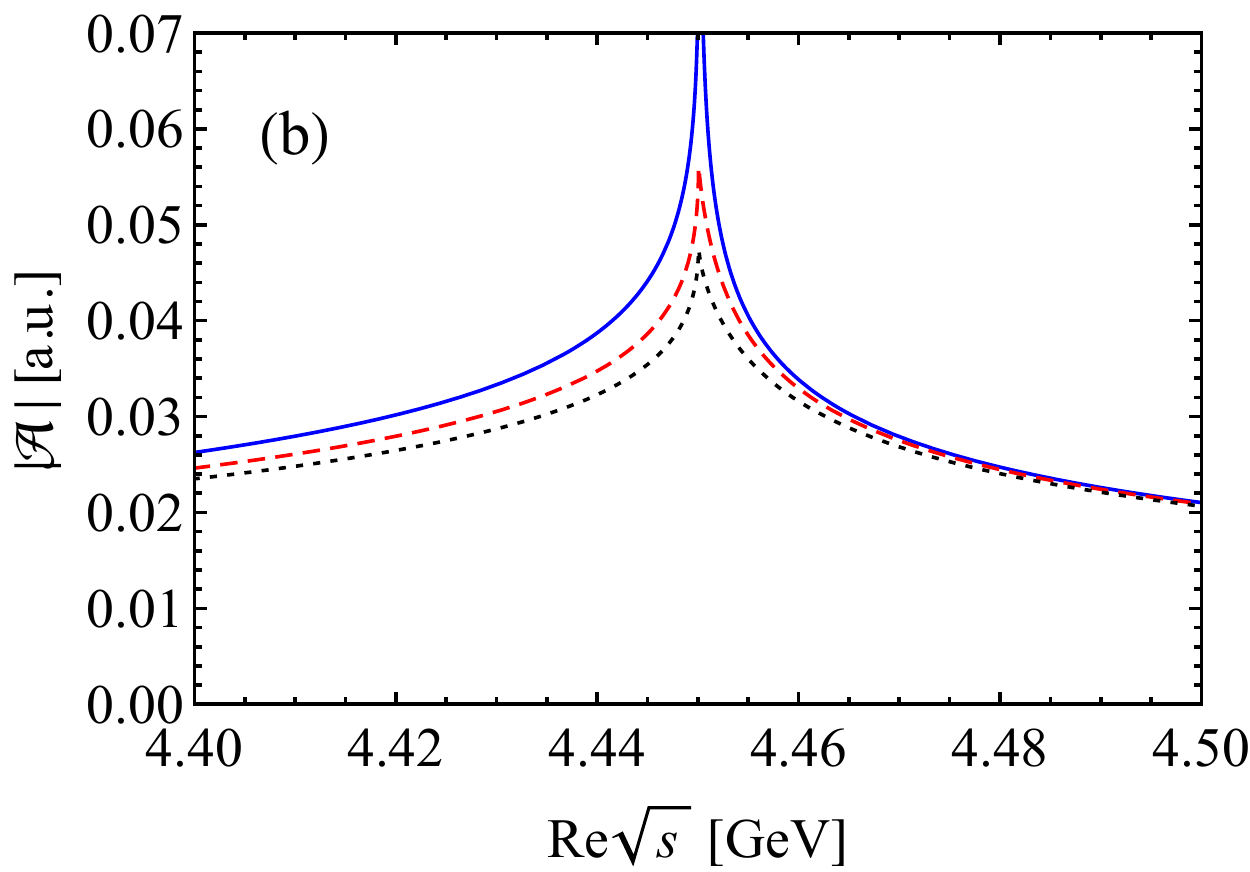}
  \caption{Absolute values of amplitudes in arbitrary units: (a) is for the
  amplitude in Eq.~\eqref{eq:Aa} fitted to the Argand plot; (b) is the for the
  triangle loop integral with the $\chi_{c1}\,p$ vertex in a $P$-wave. In (b),
  we assume the $\Lambda(1890)$ with a mass of 1.89~GeV is exchanged in the
  triangle diagram.
  The solid, dashed and dotted lines correspond to a width of the
  $\Lambda(1890)$ of 10, 60 and 100~MeV, in order.
  }
  \label{fig:amp}
\end{figure}

Using an amplitude with the loop function given in Eq.~\eqref{eq:gexplicit}, one
can get a peak around the $\chi_{c1}\,p$ threshold. In order to have a more
quantitative description of the effect of Fig.~\ref{fig:diag}~(a), we fit to the
Argand plot for the $P_c(4450)$ amplitude depicted in Fig.~9~(a) in
Ref.~\cite{Aaij:2015tga} with an amplitude
\begin{equation}
  \Amp_\text{(a)} = N\, \left[b + G_\Lambda(E)\right] ,
  \label{eq:Aa}
\end{equation}
where $b$ is a constant background term which may originate from a direct
production of the $K^- J/\psi\, p$, and $N$ is an overall normalization. We fit to
both the real and imaginary parts of the $P_c(4450)$ amplitude by minimizing the
sum of the chi-squared values for both the real and imaginary parts. The best
fit with a real background term has $\chi^2/{\rm d.o.f.}=1.75$ and is given by
$N=3144$, $b=-2.9\times10^{-4}$~GeV$^4$ and $\Lambda=0.16$~GeV. With a real
background term, the amplitude in Eq.~\eqref{eq:Aa} can only be complex when the
energy is larger than the $\chi_{c1}\,p$ threshold, as is evident in
Fig.~\ref{fig:argand}.  The background is in general complex as a
result of the fact that the $K,J/\psi$ and $p$ can go on shell and
many $\Lambda$ resonances can contribute to the $K\,p$ state. One sees from the
figure that the  counterclockwise feature of the LHCb amplitude is
reproduced, and
the overall agreement is good. The absolute value of the amplitude in
Eq.~\eqref{eq:Aa} with these determined parameters has a narrow peak around the
$\chi_{c1}\,p$ threshold as shown in Fig.~\ref{fig:amp}~(a).
%%%%%%%%%%%%%%%%%%%%%%%%%%%%%%%%%%%%%%%%%%%%%%%%%%%%%%%%%%%%%%%%%%%%%%%%%%%%%%%%
%%%%%%%%%%%%%%%%%%%%%%%%%%%%%%%%%%%%%%%%%%%%%%%%%%%%%%%%%%%%%%%%%%%%%%%%%%%%%%%%
We have checked that using a different form factor $\Lambda^4/\left( \vec q\,^2 +
\Lambda^2 \right)^2$ gives a similar result. In both cases, the peak is
asymmetric unlike the Breit--Wigner form.

There can be further enhancement around the $\chi_{c1}\,p$ threshold due to the
presence of nearby triangle singularities, also called leading Landau
singularities of a triangle diagram, from Fig.~\ref{fig:diag}~(b).
The leading Landau singularities for a triangle diagram are solutions of the
Landau equation~\cite{Landau:1959fi}
\begin{equation}
    1 + 2\, y_{12}\, y_{23}\, y_{13} = y_{12}^2 + y_{23}^2 + y_{13}^2,
    \label{eq:Landau}
\end{equation}
where $y_{ij} = (m_i^2 + m_j^2 - p_{ij}^2)/(2\,m_i\,m_j)$
with $m_i (i=1,2,3)$ masses of the intermediate particles, and  $p_{ij} = p_i +
p_j$ being the four momentum of the $ij$ pair. To be specific, we let $m_1, m_2$
and $m_3$ correspond to the masses of the $\Lambda^*$, $J/\psi$ and proton,
respectively. Then, $p_{12}^2 = M_{\Lambda_b}^2$, $p_{13}^2 = M_{K^-}^2$, and
$p_{23}^2 = s$ is the invariant mass squared of the $J/\psi\, p$ pair. It is
easy to solve this equation for any given variable. We solve it as an equation
of $s$, which has two solutions.
%%%%%%%%%%%%%%%%%%%%%%%%%%%%%%%%%%%%%%%%%%%%%%%%%%%%%%%%%%%%%%%%%%%%%%%%%%%%%%%%
\begin{figure}[t]
  \centering
    \includegraphics[height=6.8cm]{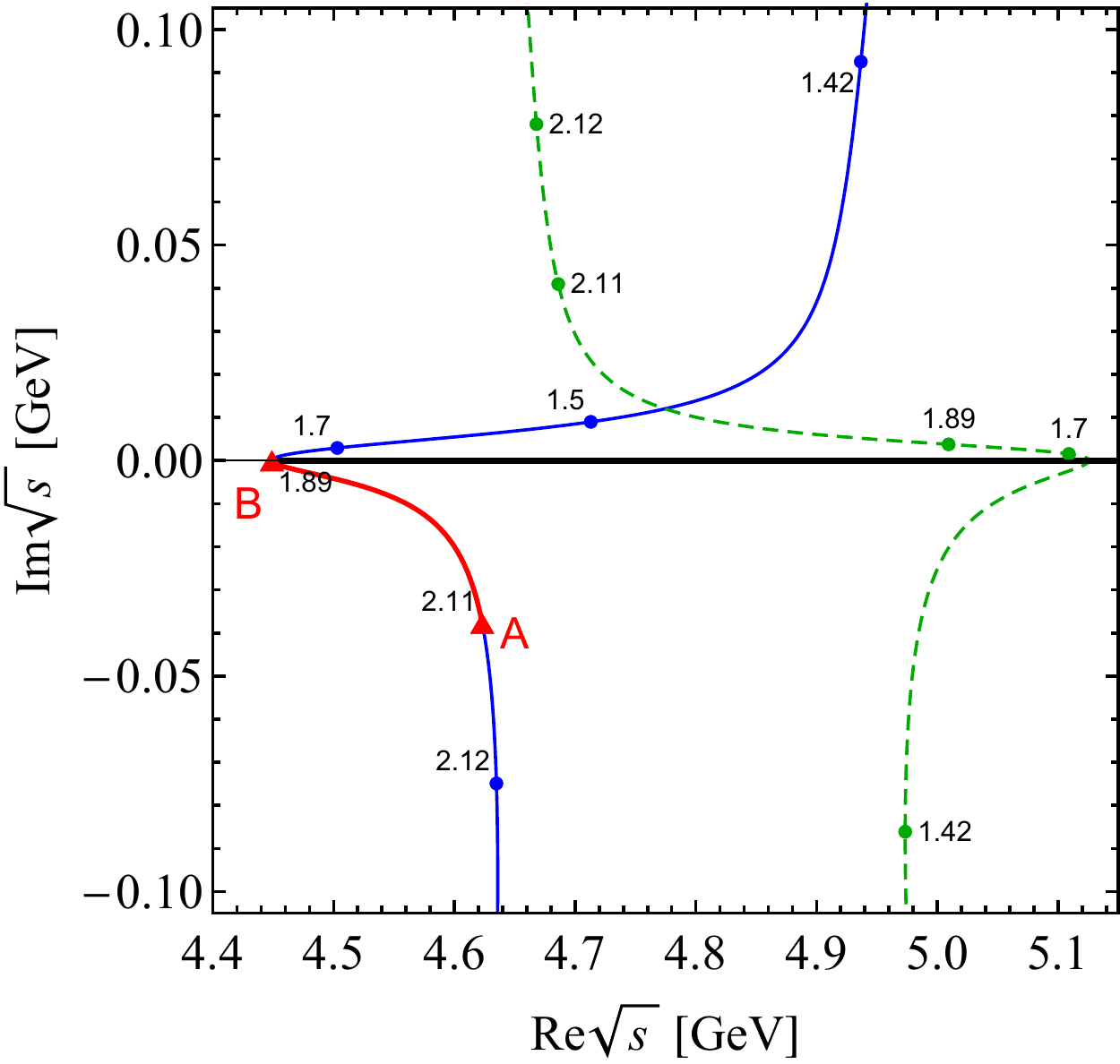} \hfill
    \includegraphics[height=6.8cm]{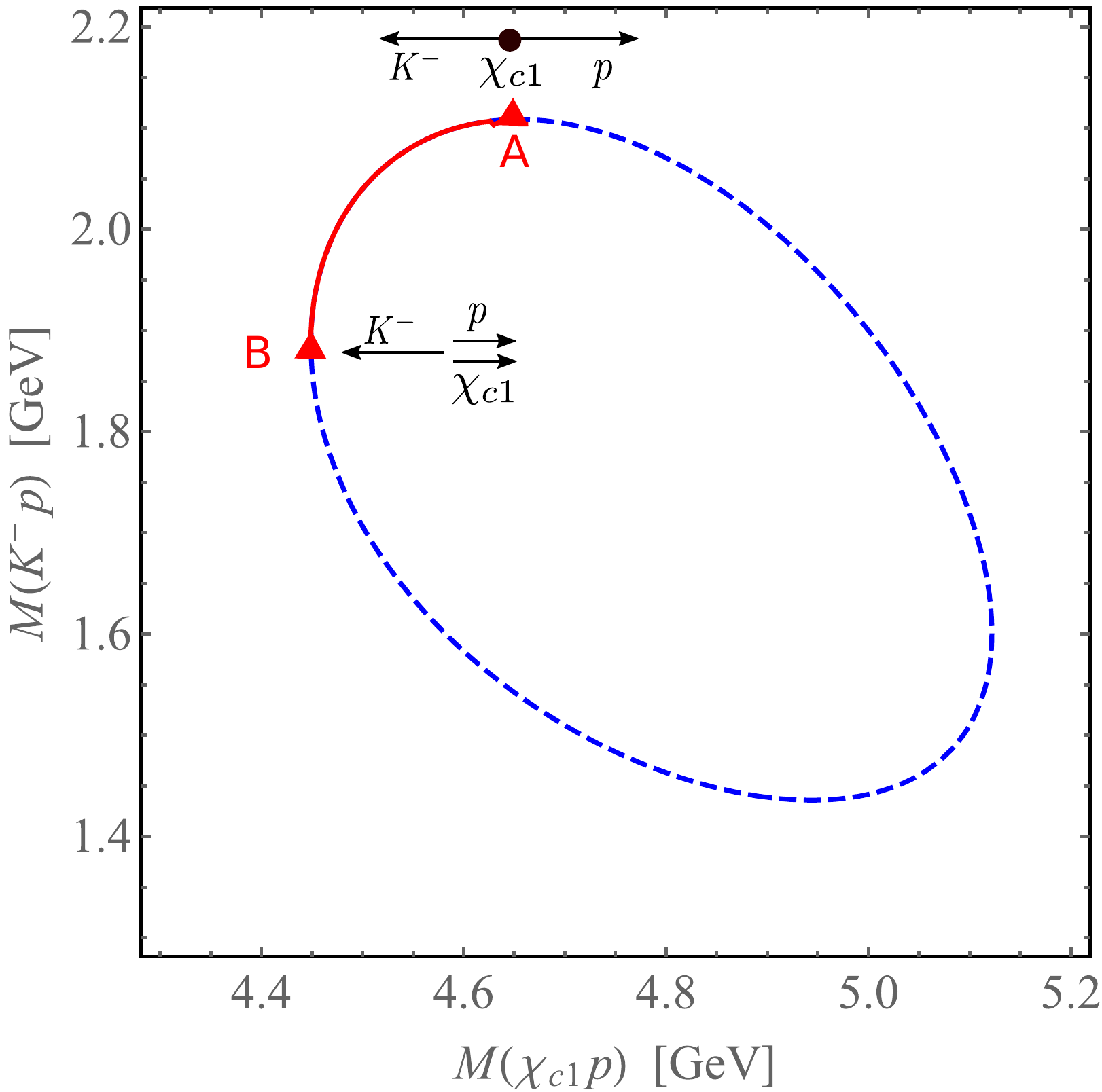}
  \caption{Left: Motion of the two triangle singularities in the complex plane
  of $\sqrt{s}=M_{\chi_{c1}p}=M_{J/\psi\,p}$ with respect to changing the mass
  of the exchanged $\Lambda^*$ baryon (several values are labeled in the plot in units of GeV). In order to
  distinguish the trajectories from the real axis, we put a small imaginary
  part, $-5$~MeV corresponding to a width of 10~MeV, to $M_{\Lambda^*}$. Only
  the part between the two filled triangles, labelled as A and B, has a large
  impact on the physical amplitude. The thick solid straight line represents the unitary cut starting
  from the $\chi_{c1}\,p$ threshold. Right: The corresponding Dalitz plot which
  shows the region between A and B. }
  \label{fig:sing}
\end{figure}
%%%%%%%%%%%%%%%%%%%%%%%%%%%%%%%%%%%%%%%%%%%%%%%%%%%%%%%%%%%%%%%%%%%%%%%%%%%%%%%%
For an easy visualization, we plot in the left panel of
Fig.~\ref{fig:sing} the motion of the solutions in the complex $\sqrt{s}$ plane.
As discussed in 1960s, see e.g. Ref.~\cite{Schmid:1967}, only one of the
singularities can have an impact on the amplitude in the physical region defined
on the upper edge of the real axis on the first Riemann sheet of the complex
$s$-plane, and it is effective only in a limited region of one of these
variables. Here we want to investigate in which values the $\Lambda^*$ mass can take
so that  there can be an evident singularity effect in the $J/\psi\, p$ invariant
mass, $\sqrt{s}$. According to the Coleman--Norton theorem~\cite{Coleman:1965xm},
the singularity is in the physical region only when the process can happen
classically, which means that all the intermediate states are on shell, and the
proton emitted from the decay of the $\Lambda^*$ moves along the same direction
as the $\chi_{c1}$ and can catch up with it to rescatter. Let us start from a
very large mass for the $\Lambda^*$ so that it cannot go on shell in
Fig.~\ref{fig:diag}~(b). Decreasing this mass, when it has a value
\begin{equation}
m_{1,\text{high}} = \sqrt{p_{12}^2} - m_2\, ,
\end{equation}
it can go on shell. At this point, the $\chi_{c1}$ is at rest in the rest frame
of the decaying particle $\Lambda_b$, and the proton emitted from the decay
$\Lambda^*\to K^- p$ can definitely rescatter with the $\chi_{c1}$ classically.
This is the point shown as a filled triangle with $M_{\Lambda^*}=2.11$~GeV,
labelled as A, on the solid curves in Fig.~\ref{fig:sing}. If we decrease $m_1$
further, the $\chi_{c1}$ will speed up and the proton will slow down. Thus, the lower bound
of $m_1$ for the rescattering process that happens classically is given by the
case when the $\chi_{c1}$ and the proton are at a relative rest, i.e. when the
$\chi_{c1}\,p$ invariant mass is equal to their threshold.
Thus, at this point the triangle singularity coincide with the normal threshold,
and one gets
\begin{equation}
 m_{1,\text{low}} = \sqrt{\frac{p_{12}^2 m_3 + p_{13}^2 m_2}{m_2
 + m_3}  - m_2 m_3 }\, .
\end{equation}
If $m_1$ is smaller than $m_{1,\text{low}}$, the proton would not be able to
catch up with the $\chi_{c1}$ and the triangle diagram can only be a quantum
process. For the case of Fig.~\ref{fig:diag}~(b), $m_{1,\text{low}}$ is given by
$M_{\Lambda^*}=1.89$~GeV, labelled as B and also shown as a filled triangle in
Fig.~\ref{fig:diag}. In the left panel of Fig.~\ref{fig:sing}, in order to move
the singularity trajectories away from the real axis, we give a 10~MeV width to the $\Lambda^*$. For a vanishing width, the solid and dashed
trajectories would pinch the real axis at $m_1 = m_{1,\text{high}}$.
We can now know on which Riemann sheet of the complex $s$-plane the
singularities are located. Since only when $m_1$ is between
$m_{1,\text{low}}$ and $m_{1,\text{high}}$ (the part between the two filled
triangles in the figure), the process can happen classically and the singularity
can be on the physical boundary (if the $\Lambda^*$ width vanishes), we conclude
that the singularity shown as the solid curve is always on the second Riemann
sheet. On the contrary, the singularity whose trajectory is shown as the dashed
curve in the left panel of Fig.~\ref{fig:sing} is on the second Riemann sheet
when it is above the real axis, and it moves into the lower half plane of the first Riemann sheet
otherwise. Thus, it is always far away from the physical boundary, and does
not have any visible impact on the physical amplitude. For an easy visualization
of the kinematical region between A and B, we show the corresponding Dalitz plot
in the right panel of Fig.~\ref{fig:sing}.

An intriguing observation for the case of interest is that within the range
between 1.89~GeV and 2.11~GeV, there is a four-star baryon $\Lambda(1890)$ with
$3/2^+$. Taking  $M_{\Lambda^*}=1.89$~GeV, the triangle singularity is just
at the $\chi_{c1}\,p$ threshold which can provide a further threshold
enhancement.~\footnote{The mechanism of enhanced threshold effect due to the
triangle singularity was recently discussed for the case of $Z_{c}$ and $Z_b$
states~\cite{Szczepaniak:2015eza}.} Giving a finite width to the $\Lambda(1890)$,
the singularity moves away from the real axis into the lower half
plane of the second Riemann sheet (it is located at $(4447-i\,0.2)$~MeV for
$M_{\Lambda^*}=(1.89-i\,0.03)$~GeV), and the enhancement is reduced.
The $\Lambda(1890)$ has a relatively small width (60 to
100~MeV~\cite{Agashe:2014kda}) so that there can still be an important enhancement.
In Fig.~\ref{fig:amp}~(b), we show the absolute value of the triangle loop
integral with the $\chi_{c1}\,p$ in a $P$-wave for three different widths (for a
discussion of the triangle singularities in nonrelativistic triangle loop
integral, see Ref.~\cite{Guo:2014qra}).
There is clearly an enhancement nearby 4.45~GeV even when the width is taken to be
100~MeV.

In the above, we have shown that kinematical effects can result in a narrow
structure around the $\chi_{c1}\,p$ threshold in the $J/\psi\, p$ invariant mass of
the $\Lambda_b^0\to K^- J/\psi\, p$ decay. Consequently, a natural question is whether
such an effect happens at other thresholds, in particular those related to the
$\chi_{c1}\,p$ through heavy quark spin symmetry (HQSS). As a result of the HQSS,
the operator for annihilating a $\chi_{c1}$ and creating a $J/\psi$ is contained
in
\begin{equation}
 \frac12\left\langle J^\dag \chi^i \right\rangle =
 -\psi^{j\,\dag}\chi_{c2}^{ij} -  \frac1{\sqrt{2}}\epsilon^{ijk}\psi^{j\,\dag}
 \chi_{c1}^k + \frac{1}{\sqrt{3}}\psi^{i\,\dag}\chi_{c0} +
 \eta_c^\dag h_{c}^i \, ,
\end{equation}
where the fields $J=\vec{\psi}\cdot\vec{\sigma}+\eta_c$ and $\vec{\chi}=\sigma^j
\left(-\chi_{c2}^{ij}-\frac{1}{\sqrt{2}}\epsilon^{ijk}\chi_{c1}^k +
\frac{1}{\sqrt{3}}\delta^{ij}\chi_{c0} \right) +
h_c^i$~\cite{Fleming:2008yn,Guo:2010ak} annihilate the $S$-wave and $P$-wave
charmonium states, respectively. This means that the rescattering interaction
strength for $\chi_{c2}\,p\to J/\psi\, p$ or $\chi_{c0}\,p\to J/\psi\, p$ is of
similar size as that for the $\chi_{c1}\,p\to J/\psi\, p$. One might naively
expect enhancements at both the $\chi_{c2}\,p$ and $\chi_{c0}\,p$ thresholds in
the $J/\psi\, p$ invariant mass as well. However, this is not the case. As we
have shown in Eq.~\eqref{eq:operators}, at  leading order in $\alpha_s$,
the charmoium  is produced by the $[\bar c\gamma^{\mu}(1-\gamma_5) c]$ current.
This current has no projection onto the $\chi_{c0}$ or $\chi_{c2}$. The
production of the $\chi_{c0,c2}$ in the $b$ decays can  come only from
higher-order QCD corrections which are  suppressed. Indeed, there is no
enhancement at the $\chi_{c2}\,p$ and $\chi_{c0}\,p$ thresholds in $\Lambda_b$
decays, which is consistent with our expectation.

The above analysis is applicable to any $b$ quark decay in which the
$\chi_{c0,c2}$ is directly generated  by the weak interaction. But it would
be different if the initial decay heavy particle contains a charm or anticharm
quark in addition to the bottom quark.  Processes of this type  include the
decays of the $B_c$ meson and the doubly-heavy baryon $\Xi_{bc}$. An explicit
calculation of $B_c$ decays \cite{Wang:2009mi} indicates that the $\chi_{c0,c2}$
can have similar production rates with the $\chi_{c1}$.
Considering the large amount of  data on the $B_c$ to be accumulated  by the LHCb
collaboration~\cite{Bediaga:2012py},  it appears very promising
to investigate the $\chi_{c0}\,p$ and $\chi_{c2}\,p$ threshold effects in the
future.
In addition, one can study the threshold effects in the prompt production of
the  $J/\psi\, p$ at the LHC, or in the $\Upsilon(1S)$ decays into the
$\chi_{cJ}\, p\, \bar p$ and $J/\psi\, p\, \bar p$.

In conclusion, what we have shown here is that the present information
on the narrow structure around 4.45~GeV observed by the LHCb Collaboration is
compatible with kinematical effects around the $\chi_{c1}\,p$ threshold: First,
it is located exactly at the  $\chi_{c1}\,p$ threshold. Second, the mass of the
four-star well-established $\Lambda(1890)$ coincidentally makes the triangle
singularity on the physical boundary located at the $\chi_{c1}\,p$ threshold,
despite a small shift into the complex plane due to the finite width of the
$\Lambda(1890)$, and third, the $\chi_{c1}$, instead of the $\chi_{c0}$ or
$\chi_{c2}$, can be easily produced in the weak decays of the $\Lambda_b$ by the
$V-A$ current so that there can be an evident effect at the $\chi_{c1}\,p$,
but not the $\chi_{c0}\,p$ or $\chi_{c2}\,p$, threshold.

Therefore, the most important question regarding the structure around 4.45~GeV
is whether it is just a kinematical effect or a real resonance. As discussed in
Ref.~\cite{Guo:2014iya}, kinematical singularities, including both the normal
threshold and the triangle singularity, cannot produce a narrow near-threshold peak
in the elastic channel, which is the $\chi_{c1}\,p$ in this case. The reason is
the interaction strength in the elastic channel controls the threshold behavior,
and there can be a narrow near-threshold peak only when the interaction in the
elastic channel is strong enough to produce a pole in the $S$-matrix which
corresponds to a real resonance. On the contrary, one cannot simply determine
the interaction strength for the inelastic channel ($\chi_{c1}\,p\to J/\psi\, p$
in our case) because it can always interfere with a direct production of the final
state. Thus, the question can be answered by analyzing the process
$\Lambda_b^0\to K^-\chi_{c1}\,p$: if there is a narrow structure just above
threshold in the $\chi_{c1}\,p$ invariant mass distribution, then the structure
cannot be just a kinematical effect and calls for the existence of a real
pentaquark-like exotic resonance, otherwise, one cannot conclude the $P_c(4450)$
to be another exotic hadron.

\medskip

\section*{Acknowledgments}

This work is supported in part
by DFG and NSFC through funds provided to the Sino-German CRC 110
``Symmetries and the Emergence of Structure in QCD'' (NSFC Grant No. 11261130311)
and by NSFC (Grant No.~11165005), by the Chinese Academy of Sciences (CAS)
President's International Fellowship Initiative (PIFI) (Grant No.
2015VMA076), by Shanghai Natural  Science Foundation  under
Grant  No. 11DZ2260700 and No. 15ZR1423100,  by the Open Project Program of
State Key Laboratory of Theoretical Physics, Institute of Theoretical Physics,
Chinese  Academy of Sciences, China (No.Y5KF111CJ1), and  by the Scientific
Research Foundation for the Returned Overseas Chinese Scholars, State
Education Ministry.

\end{document}